\documentclass{article}

\makeatletter
\newcommand{\chapterauthor}[1]{%
 {\parindent0pt\vspace*{0pt}%
 \linespread{1.0}\normalsize\scshape#1%
 \par\nobreak\vspace*{15pt}}
 \@afterheading%
}
\makeatother

\usepackage[a4paper,margin=1in]{geometry}
\usepackage{parcolumns,lipsum}
\usepackage{verbatim}
\usepackage{subcaption}
\usepackage{url}
\usepackage{xcolor}
\usepackage{changepage}

\title{Slow neutrons in Palermo:\\a forgotten conference by Enrico Fermi}
\author{Emanuele Goldoni, Ledo Stefanini\\ \textit{Accademia Nazionale Virgiliana}\\ \textit{Mantova, Italy}}
\begin{document}
\maketitle

\begin{abstract}
On October 22, 1934, in a famous experiment, Enrico Fermi and his colleagues discovered that a significant increase in induced radioactivity can be obtained when neutrons are slowed down by means of hydrogen atoms. This discovery and its explanation earned him the 1938 Nobel Prize in Physics.
One year later, on October 1935, Fermi held a public speech in Palermo, Italy, presenting his findings at the 24th congress of the Italian Society for the Progress of Sciences. The transcription of his speech, entitled ``Recenti risultati della radioattività artificiale'' (Recent Results in Artificial Radioactivity), was soon published in Italy. The published text is one of the very first articles in which Fermi describes the success of the Via Panisperna boys and how he came to discover that hydrogen atoms slowed the neutrons.
Unfortunately, the content of the original Italian speech has never been republished since and was forgotten. Similarly, a translation published internally by the US Atomic Energy Commission is not publicly available and is probably lost. In this work, we include the transcription of the original article in Italian, and we also propose an English translation to make the text available worldwide to a broader public.
\end{abstract}

\section*{Introduction}

The 1938 Nobel Prize in Physics was awarded to Enrico Fermi ``for his demonstrations of the existence of new radioactive elements produced by neutron irradiation, and for his related discovery of nuclear reactions brought about by slow neutrons'' \cite{Nobel:1938:outreach}.

The history of this discovery started in January 1934, when Irene Joliot-Curie and her husband Frédéric discovered that radioactivity can also be induced in stable elements by bombarding them with $alpha$ particles. Using particles emitted from a polonium source, they transformed aluminum, boron, and magnesium nuclei into radioactive isotopes. 

Upon hearing the news of the discovery of artificial radioactivity, Fermi had the idea of trying to produce new radioactive elements using neutron sources instead of the $alpha$ particles used by the French researches. What he needed was just a neutron source, and he knew how to build a high-intensity one. The process involved collecting radon gas (an emanation of radium) in a test tube, immersing it in liquid nitrogen so that the gas became liquid, adding beryllium powder, and welding everything together. On March 20, 1934, Fermi obtained induced radioactivity for the first time through neutron bombardment. From a laboratory notebook that was believed to have been lost -- and, luckly, found in 2002 -- we know that the research began with platinum and then extended to the entire periodic table \cite{Guerra:2005:quaderno}.

After the initial results, Fermi and his collaborators quickly published their new experimental findings in a series of letters to {\em La ricerca scientifica}, the Italian National Research Council (CNR) journal. The first article was dated on 25 March 1934 \cite{FermiE:1934:radioattivita}, just a few days after the first result: this also attests to the urgency of the matter. In a few months, the Roman team irradiated over sixty elements with neutrons producing forty-four new isotopes by artificial radioactivity. However, after returning from summer vacation, things started to go wrong. Their whole experimental activity was affected by a strange inconsistency in the results of irradiation of the targets. 
Apparently, the readings turned out to depend on the tables used as a support of the equipment. One table was made of wood; the other, not far away, was made of marble. When the experiment on inducing radioactivity in silver samples was carried out on the wooden table, a significantly greater activity resulted than if the same experiment was tried on the stone support \cite{Holton:2005:tables}.

To make sense of the results, in October 1934 Fermi and his team reasoned that the interposition of a block of lead between the source and the target changed somewhat the activation. Therefore, Fermi decided to make a lead filter, a wedge of varying thickness, to insert into the neutron beam. 
Then, on 22 October 1934, something strange happened. As Fermi recalled to Subrahmanyan Chandrasekhar \cite{Chandrasekhar:1962:how}:

\begin{adjustwidth}{2em}{2em}{\em One day, as I came to the laboratory, it occurred to me that I should examine the effect of placing a piece of lead before the incident neutrons. And instead of my usual custom, I took great pains to have the piece of lead precisely machined. I was clearly dissatisfied with something: I tried every "excuse" to postpone putting the piece of lead in its place. When finally, with some reluctance, I was going to put it in its place, I said to myself: `No! I do not want this piece of lead here; what I want is a piece of paraffin.' It was just like that: with no advanced warning, no conscious, prior, reasoning. I immediately took some odd piece of paraffin I could put my hands on and placed it where the piece of lead was to have been.}
\end{adjustwidth}

The result was immediately obvious: a great increase in the radioactivity induced in the target, even if the target and the paraffin filter were placed on the stone shelf. As Segrè recalled \cite{Segre:1970:physicist}, at about noon everyone was summoned to watch the miraculous effect of paraffin filtration. They went for lunch and when they came back, at about three in the afternoon, Fermi had found the explanation of the strange behavior of filtered neutrons. He hypothesized that neutrons could be slowed down by elastic collisions, and in this way become more effective; an idea that was contrary to the expectations.

Fermi's great merit was his intuition in explaining the phenomenon: a discovery that he later described as ``{\em the most important I have made}''. Since the neutron has the same mass as the proton, in the collision between the projectile neutron and the stationary proton, the neutron transfers part of its kinetic energy. After about 20 collisions, the energy of the neutron is reduced to thermal energy. Now, the cross section between a heavy nucleus and a proton is much higher for slow neutrons than for fast ones, which explains their greater effectiveness in nuclear reactions with heavy nuclei. 

On October 12, 1935 Fermi held a public speech in Palermo, Italy, at the 24th congress of the SIPS - Società Italiana per il Progresso delle Scienze (Italian Society for the Progress of Sciences) presenting his discovery. The text of his speech, entitled ``Recenti risultati della radioattivita artificiale'' (Recent Results in Artificial Radioactivity), was soon published in Italy \cite{Fermi:1935:recenti}, and then translated into English by the US Atomic Energy Commission and issued as an internal document \cite{Fermi:1935:recents}. This text was one of the very first articles published by Fermi which describes the success of the via Panisperna boys and how he came to discover that hydrogen atoms slowed the neutrons. Notably, some passages of the text - mainly the description of the experiment with the rotating wheel - are similar to those used by Fermi in the Nobel Lecture he held three years later in Stockholm. In addition, in the short article we present here, Fermi uses an extraordinarily effective comparison between paraffin and ``a kind of solution of slow neutrons''. This image reflects the first idea that, seven years later in Chicago, led him from the idea of the chain fission reaction to the creation of the graphite-moderated atomic pile.

Unfortunately, the original Italian text of the speech was not included in the well-known {\em Collected Papers} \cite{Fermi:1962:collected1} and it has never been republished since Fermi's death. Similarly, the original translation is not publicly available and is probably lost or still buried in the archives. In this article, we include the transcription of the original Italian speech, and we also propose an English translation to make the text available worldwide and accessible to a broader public.

\section*{Recenti risultati della radioattività artificiale}

\chapterauthor{Enrico Fermi}

Mi limiterò ad un rapido riassunto di quanto è stato fatto l'anno scorso nell'Istituto Fisico di Roma sulla radioattività provocata da bombardamento di neutroni. Questo è un caso particolare della radioattività artificiale il cui primo esempio fu scoperto due anni or sono da J. Curie-Joliot e F. Joliot.

È noto che in molti casi quando un nucleo è colpito da una particella esso si trasmuta in un nuovo nucleo. In tale processo la particella proiettile rimane incorporata nel nuovo nucleo. La durata del processo di trasmutazione può essere valutata dall'ordine di $10^{-20}$ secondi. Se il nuovo nucleo è instabile, esso si disintegra; questo fenomeno prende il nome di radioattività artificiale. I corpuscoli urtanti usati nelle ricerche di cui debbo trattare sono neutroni. La massa di questi corpuscoli è $1$ cioè uguale all'unità elementare dei pesi atomici; e la loro carica è zero. Essi si prestano assai meglio delle particelle $alpha$ a produrre radioattività artificiale perché queste ultime sono corpuscoli carichi positivamente e quindi sono respinti dalla carica positiva del nucleo; naturalmente questo ostacolo non si presenta per i neutroni. Nelle esperienze eseguite a Roma, si sono provati tutti gli elementi possibili (circa 80) e il 60 percento presentano il fenomeno della radioattività artificiale. Le ricerche eseguite hanno anzi mostrato che esistono tre diversi processi possibili per la formazione di nuclei radioattivi.

Tutti e tre questi processi cominciano con la cattura del neutrone da parte del nucleo bombardato, cattura che avviene in un tempo di circa $10^{-20}$ secondi. Nello stesso tempo viene emessa o una particella alfa o un protone o nulla.

Poiché la particella $alpha$ ha una carica $+2$ ed il protone una carica $+1$, il numero atomico $Z$ nel primo processo diventa $Z-2$, nel secondo $Z-1$, nel terzo rimane invariato. Nei due primi casi l'elemento si trasforma in un nuovo elemento chimico spostato rispettivamente di due od un posto nel sistema periodico di Mendelejeff, mentre nel terzo caso si ha semplicemente la formazione di un nuovo isotopo perché la massa è cresciuta di una unità. Le ricerche per stabilire secondo quale di questi tre schemi avviene un determinato processo si fanno per via chimica,

Le ricerche eseguite sistematicamente ci hanno permesso di stabilire il meccanismo di tutte le reazioni. Non vorrei fermarmi sopra i dettagli di tali processi ma accennare allo sviluppo più recente dei lavori.

A priori si potrebbe pensare che più forte è un urto e maggiore è l'efficacia dei neutroni nel produrre nuovi nuclei. Ci sono invece dei casi in cui conviene ridurre la velocità delle particelle urtanti; di questo fatto ci siamo accorti in modo casuale. In tutte queste ricerche la sorgente dei neutroni è costituita da un aghetto di vetro contenente emanazione e polvere di Berillio; da essa esce circa un milione di neutroni al secondo. La sostanza da irradiare viene posta accanto alla sorgente. Noi ritenevamo che il numero di urti dei neutroni contro i nuclei fosse proporzionale all'angolo solido e quindi inversamente proporzionale al quadrato della distanza. Nel corso di alcune esperienze notammo che delle anomalie a tale riguardo. E precisamente osservammo che dei pezzi di piombo posti nelle vicinanze disturbavano il fenomeno nel senso di aumentarne l'effetto; la cosa riusciva a prima vista incomprensibile.

Esperienze di questo genere sono state allora eseguite con altre sostanze. Si provò a porre dei blocchetti di paraffina fra la sorgente e l'oggetto da irradiare. Più precisamente l'esperienza venne fatta nel modo seguente. Si misurava quanto diventava radioattiva una sostanza una volta interponendo ed una volta non interponendo la paraffina. Contrariamente a quanto credevamo si trovò una radioattività di circa il 40\% maggiore quando vi era la paraffina. Si circondò allora tutto di paraffina e si trovò che in queste condizioni l'effetto si moltiplicava per 40 0 50.

Vediamo ora di spiegare questo fenomeno. Siccome la paraffina contiene $C$ e $H$ si trattava in primo luogo di stabilire a quale di questi due elementi si doveva attribuire l'effetto osservato. Ripetendo l'esperienza con altre sostanze diverse dalla paraffina fu possibile stabilire che l'effetto è dovuto all'$H$.

La particolarità di questo elemento è di essere il più leggero di tutti; si noti anzi che esso ha massa uguale al neutrone.

Vediamo ora cosa accade quando un neutrone urta un atomo; se questo è molto pesante, per esempio se ha massa atomica 100, il neutrone non lo sposta e rimbalza conservando tutta la sua energia. Se l'urto avviene invece contro un atomo di idrogeno ossia contro un corpo di eguale massa, il comportamento è tutto differente. Nell'urto il neutrone rimbalza ma trasmette un impulso notevole all'atomo urtato. Ne segue che in un tale urto l'energia del neutrone si riduce molto. Si dimostra anzi facilmente che in media l'energia si divide per il fattore $e = 2,71$. Con una ventina di urti successivi l'energia si ridurrà quindi di un fattore $10^{4}$. Alla fine di tale processo il neutrone sarà dotato di un'energia piccolissima diversa da zero perché gli atomi di H si muovono per l'energia termica. I neutroni si trovano allora nella paraffina soggetti all'agitazione termica. Questi proiettili che escono dalla sorgente con un'energia di $4 \times 10^{4}$ Volt sono ben presto ridotti a particelle che si muovono con un'energia di solo $1/30$ di Volt..

La paraffina si presenta così come una specie di soluzione di neutroni lenti.

Uno studio sistematico dell'azione dei neutroni lenti nel provocare sostanze radioattive artificiali ha mostrato che essi sono particolarmente efficaci in tutti quei processi in cui si ha semplice cattura del proiettile urtante mentre sono incapaci di dar luogo a reazioni nucleari in cui si abbia la emissione di una particella $alpha$ o di un protone.

Per controllare la correttezza della interpretazione data è stata fatta un'altra esperienza. Si trattava di controllare che questi neutroni si diffondono nell'ambiente con una velocità corrispondente alla agitazione termica e quindi dipendente dalla temperatura. Una esperienza diretta a tale scopo diede risultato negativo. Oggi credo di sapere il perché di tale risultato. Poco dopo l'effetto fu trovato da altri (Moon e Tillman) che fecero una analoga esperienza con un dispositivo analogo. In un vaso Dewar veniva posto un blocchetto di paraffina che poteva essere raffreddato con l'aria liquida. Attorno al vaso veniva posta altra paraffina e in questa la sorgente. La sostanza da attivare era posta entro il vaso di Dewar. L'esperienza ha confermato che la temperatura esercita una influenza nell'attivazione dei varii elementi. La spiegazione data è quindi corretta.

Per concludere accennerò ad un'altra esperienza fatta sempre per mettere in evidenza il meccanismo di questo fenomeno. Lo scopo era di determinare per quanto tempo i neutroni lenti restassero in soluzione nella paraffina prima di venire catturati dagli atomi di idrogeno e scomparire dando luogo alla formazione di un deutone. Sostanzialmente il problema è questo: noi abbiamo una sorgente che emette neutroni, i quali si muovono attraverso la paraffina, subiscono degli urti, e vengono di conseguenza rallentati; essi seguitano a girare ma non indefinitamente perché dopo un po' finiscono catturati da un protone. Si tratta di vedere per quanto tempo in media restano nell'ambiente. Si intende che il tempo che ci restano è molto breve ma pur tuttavia assai lungo rispetto al tempo che impiega un neutrone veloce che esce dalla sorgente a percorrere una distanza dell'ordine di grandezza del percorso di un neutrone lento. L'esperienza è stata eseguita al modo seguente: abbiamo preso una grande ruota d'acciaio che veniva fatta girare da un motore a velocità abbastanza grande (la velocità periferica era di 60 metri al secondo; se fosse stato necessario si sarebbe potuta ottenere una velocità più grande). Sopra il bordo di questa ruota era fissato un piccolo astuccio di acciaio in cui si poneva la sorgente di neutroni. Dalle due parti di questo astuccio erano fissate pure al bordo della ruota due lastrine di manganese che è una sostanza in cui si osserva assai bene il fenomeno della radioattività artificiale; queste due lastrine servivano da rivelatori dei neutroni, cioè con la loro attività indicavano la presenza di neutroni. Da una parte e dall'altra della ruota e l'una simmetrica all'altra c'erano due ciambelle di paraffina fisse ed assai vicine alla ruota che girava tra di loro. In un primo momento pensiamo di tenere la ruota fissa, di porre la sorgente nel suo astuccio e di lasciarcela per esempio per due ore; poi immaginiamo di togliere la sorgente, prendere le due lastrine, portarle ad un elettrometro e misurarne l'attività. È abbastanza chiaro che se si fa l'esperienza in queste condizioni si deve trovare eguale l'attività nelle due lastrine per evidenti ragioni di simmetria. Ora supponiamo di ripetere l'esperienza mettendo in rapida rotazione la ruota. Allora vediamo quello che ci si deve attendere: pensiamo di seguire la sorte dei neutroni che vengono emessi dalla sorgente in un certo intervallo di tempo; tutti quelli che vengono emessi negli istanti successivi subiranno la medesima sorte. Se la ruota fosse ferma questi neutroni escono dalla sorgente in tutte le direzioni e la maggior parte di essi va a finire nella paraffina ove subiscono vari urti, vengono rallentati e diffusi un po' in tutte le direzioni; alcuni di essi riescono e vanno a colpire le due lastrine di manganese attivandole. Ora se la ruota gira ed i neutroni in media restano qualche tempo entro la paraffina quelli che ne riescono troveranno la ruota un pochino spostata rispetto alla posizione che aveva nello istante in cui erano stati emessi dalla sorgente; ne segue che la lastrina che si trova sul didietro della sorgente verrà colpita da più neutroni di quella che è davanti. Dovremo quindi attenderci che la lastrina posteriore venga attivata con maggiore intensità che non la lastrina anteriore. L'esperienza ha confermato questa previsione ed ha permesso di stabilire che in media i neutroni restano $10^{-4}$ secondi entro la paraffina prima di venire catturati dagli atomi di idrogeno.

Aggiungerei ancora due parole sulla possibilità che i fenomeni della radioattività acquistino interesse pratico.

L'applicazione pratica più ovvia potrebbe aversi nel campo medico, un'altra applicazione si può fare in chimica analitica usando le sostanze radioattive artificiali come indicatori per seguire il comportamento delle varie sostanze. Per applicazioni di questo genere e forse per qualche altra è essenziale poter produrre queste sostanze radioattive in quantità maggiore di quanto non si sia potuto fare finora. A mezzo delle sostanze idrogenate come la paraffina e l'acqua è stato possibile accrescere di un fattore 10 o 100 l'intensità dei fenomeni di radioattività. Ma per arrivare a produrre le sostanze radioattive artificiali in quantità dell'ordine di quelle delle sostanze radioattive naturali sarà evidentemente necessario aumentare l'intensità delle sorgenti di neutroni. Per raggiungere tale scopo bisognerà sviluppare la tecnica delle cosidette sorgenti artificiali di neutroni. Già oggi sono note le varie reazioni nucleari che determinano l'espulsione dai nuclei di neutroni. Per esempio una reazione che si presenta tra le più promettenti è il bombardamento dell'idrogeno pesante con altro idrogeno di massa due. In questo processo si ha formazione di elio di massa 3 ed espulsione di neutroni. Ho citato questa reazione particolare perché essa presenta il vantaggio di avvenire ad un voltaggio non troppo elevato (100.000 – 200.000 volt). Probabilmente in un non lontano avvenire la sorgente di neutroni sarà costituita da una sorgente di ioni di idrogeno pesante, i quali vengono accelerati attraverso un campo di 200.000 volt e poi vanno ad urtare contro una sostanza contenente pure idrogeno pesante. Bisogna anzi dire che sorgenti di questo tipo sono già state fatte e funzionano abbastanza bene; in genere hanno il difetto di funzionare per un tempo brevissimo e poi per una ragione qualsiasi distruggersi.

Le difficoltà sono puramente tecniche e potranno essere superate. Quindi si può abbastanza ragionevolmente prevedere la possibilità, anche con altre reazioni un pochino dello stesso tipo, di arrivare a produrre sostanze radioattive in quantità pressoché equivalenti alle quantità che si usano per esempio nella pratica medica. E con ciò ho sostanzialmente finito. Per riassumere restano due tipo di problemi piuttosto diversi. Da un lato ricerche rivolte ad applicazioni pratiche della radioattività artificiale, tendenti soprattutto ad aumentare l'intensità delle sorgenti dei neutroni e quindi a produrre quantità considerevoli di sostanze radioattive, dall'altro lato resta lo studio del fenomeno in se, al fine di chiarirne le caratteristiche essenziali.

\section*{Recent Results of Artificial Radioactivity}

\chapterauthor{Enrico Fermi (Author), Emanuele Goldoni and Ledo Stefanini (Translators)}

I will limit myself to a quick summary of what has been done in the last year at the Istituto Fisico in Rome on radioactivity produced by neutron bombardment. This is a particular case of artificial radioactivity, of which the first example was discovered two years ago by J. Curie-Joliot and F. Joliot.

It is known that, in many cases, a nucleus is transmuted into a new nucleus when it is struck by a particle. In this process, the projectile particle remains captured in the new nucleus. The duration of the transmutation process can be estimated to be in the order of $10^{-20}$ seconds. If the new nucleus is unstable, it disintegrates; this phenomenon is called artificial radioactivity. The colliding particles used in the research I am about to discuss are neutrons. The mass of these particles is $1$, that is, equal to the elementary unit of atomic weight; and their charge is zero. Neutrons are much better than $alpha$-particles for producing artificial radioactivity because the latter are positively charged and are therefore repelled by the positive charge of the nucleus; of course, this is not an obstacle for neutrons. In the experiments carried out in Rome, all possible elements (about 80) have been tested, and 60\% of them exhibit the artificial radioactivity phenomenon. The research carried out has shown that there are three different possible processes for the formation of radioactive nuclei.

All three of these processes begin with the capture of a neutron by the bombarded nucleus, which occurs in about $10^{-20}$ seconds. At the same time, an alpha particle or a proton is emitted, or nothing at all.

Since the $alpha$ particle has a charge of $+2$ and the proton a charge of $+1$, the atomic number Z in the first process becomes $Z-2$, in the second $Z-1$, in the third it remains unchanged. In the first two cases, the element transforms into a new chemical element, moving up two or one place, respectively, in Mendeleev's periodic system, while in the third case, a new isotope is formed because the mass has increased by one unit. The investigation to establish which of these three schemes a particular process follows is done chemically.

Systematic researches have allowed us to establish the mechanism of all the reactions. I don't want to dwell on the details of these processes, but rather mention the most recent developments.

A priori, one might think that the stronger the impact, the greater the effectiveness of neutrons in producing new nuclei. Instead, there are cases in which it is better to reduce the speed of the colliding particles; we realized this by chance. In all these researches, the neutron source is a small glass bulb containing beryllium powder and radon; it emits about one million neutrons per second. The substance to be irradiated is placed next to the source. We believed that the number of times the neutrons collided with the nuclei was proportional to the solid angle and, therefore, inversely proportional to the square of the distance. During some experiments we noticed some anomalies in this regard. Precisely, we observed that some pieces of lead placed nearby disturbed the phenomenon, increasing its effect; this was incomprehensible at first glance.

Experiments of this kind were then carried out with other substances. Attempts were made placing small blocks of paraffin between the source and the object to be irradiated. More precisely, the experiment was carried out in the following way. The level of radioactivity of a substance was measured both with and without paraffin. Contrary to what we believed, the level of radioactivity was found to be about 40\% higher when paraffin was present. When everything was then surrounded by paraffin, it was found that the effect was multiplied by 40 or 50.

Let's try to explain this phenomenon. As paraffin contains $C$ and $H$, the first thing to do was to establish which of these two elements was responsible for the observed effect. By repeating the experiment with substances other than paraffin, it was possible to establish that the effect is due to $H$.

The particularity of this element is that it is the lightest of all; in fact, it has the same mass as a neutron.

Let's see what happens when a neutron hits an atom; if the atom is very heavy, for example if it has an atomic mass of 100, the neutron does not move it and bounces back conserving all its energy. If the collision occurs with a hydrogen atom, i.e. with a body of equal mass, the behavior is completely different. In the collision, the neutron bounces but transmits a considerable impulse to the hit atom. It follows that, in such a collision, the energy of the neutron is greatly reduced. In fact, it can be easily demonstrated that, on average, the energy is divided by the factor $e = 2.71$. With about twenty successive collisions, the energy will therefore be reduced by a factor of $10^{4}$. At the end of this process, the neutron will have a very small energy, different from zero because the H atoms move due to thermal energy. Hence, the neutrons are in the paraffin subject to thermal agitation. These projectiles that leave the source with an energy of $4 \times 10^{4}$ Volts are soon reduced to particles that move with an energy of just $1/30$th of a Volt.

The paraffin thus presents itself as a kind of solution of slow neutrons.

A systematic study of the impact of slow neutrons in producing artificial radioactive substances has shown that they are particularly effective in all those processes in which there is simple capture of the colliding projectile, while they are incapable of giving rise to nuclear reactions in which there is emission of an $alpha$ particle or a proton.

Another experiment was carried out to check the correctness of the given interpretation. The goal was to check that these neutrons spread in the environment at a speed corresponding to thermal agitation and, therefore, dependent on temperature. An experiment carried out for this purpose gave a negative result. Today I think I know why this happened. Shortly afterwards, the effect was found by others (Moon and Tillman) who carried out a similar experiment with a similar device. A block of paraffin was placed in a Dewar flask and cooled with liquid air. More paraffin was placed around the flask and the source was placed inside it. The substance to be activated was placed inside the Dewar flask. The experiment confirmed that temperature has an influence on the activation of the various elements. The given explanation is therefore correct.

In conclusion, I will mention another experiment that was carried out to highlight the mechanism of this phenomenon. The aim was to establish how long the slow neutrons remained in solution in paraffin before being captured by hydrogen atoms and disappear, giving rise to the formation of a deuteron. Basically the problem is the following. We have a source that emits neutrons. These neutrons move through the paraffin, suffer collisions, and are consequently slowed down; they continue to move but not indefinitely because, after a while, they end up being captured by a proton. It's a matter of seeing how long, on average, they remain in the environment. It is understood that the time they remain there is very short, but nevertheless quite long compared to the time it takes for a fast neutron that leaves the source to travel a distance in the order of magnitude of the path of a slow neutron. The experiment was carried out as follows: we took a large steel wheel that was turning at a fairly high speed thanks to a motor (the peripheral speed was 60 meters per second; if necessary, a higher speed could have been obtained). We fastened on the edge of this wheel a small steel case with the neutron source put inside. On both sides of this case, two small plates of manganese were fixed to the same edge of the wheel. Manganese is a substance in which the phenomenon of artificial radioactivity can be observed very clearly; these two plates acted as neutron detectors, i.e. their activity indicated the presence of neutrons. On both sides of the wheel, and symmetrical to each other, there were two paraffin donuts, fixed and very close to the wheel that rotated between them. At first, we think of keeping the wheel fixed, placing the source in its case and leaving it there, let's say for two hours; then we imagine removing the source, taking the two plates, bringing them to an electrometer and measuring their activity. It's quite clear that, if the experiment is carried out under these conditions, the activity in the two plates should be the same for obvious symmetry  reasons. Now, let's suppose we repeat the experiment with the wheel rotating rapidly. Let's see what we can expect: let's consider the fate of the neutrons that are emitted by the source in a certain time interval. All those that are emitted in the following instants will suffer the same fate. If the wheel were stationary, these neutrons would leave the source in all directions and most of them would end up in the paraffin, where they would suffer various collisions, be slowed down and spread in all directions; some of them would succeed and hit the two manganese plates, activating them. Now, if the wheel turns and the neutrons on average remain in the paraffin for some time, those that finally exit will find the wheel a little displaced with respect to the position it had at the moment they were emitted from the source; it follows that the plate behind the source will be hit by more neutrons than the one in front. We should therefore expect the back plate to be activated with greater intensity than the front plate. This prediction has been confirmed by experiment, which has shown that, on average, neutrons remain in the paraffin for $10^{-4}$ seconds before being captured by the hydrogen atoms.

I would like to add a few words about the possibility that the phenomena of radioactivity may acquire practical interest.

The most obvious practical application could be in the medical field; another application could be in analytical chemistry using artificial radioactive substances as indicators to follow the behavior of various substances. For applications of this kind, and perhaps for some others, it is essential to be able to produce these radioactive substances in greater quantities than has been possible up to now. By using hydrogenated substances, such as paraffin and water, it has been possible to increase the intensity of the radioactivity phenomena by a factor of 10 or 100. But in order to produce artificial radioactive substances in quantities similar to those of natural radioactive substances, it will be clearly necessary to increase the intensity of the neutron sources. To achieve this goal, it will be necessary to develop the technique of the so-called artificial neutron sources. The various nuclear reactions that cause the expulsion of neutrons from nuclei are already known. For example, one of the most promising reactions is the bombardment of heavy hydrogen with other mass-2 hydrogen. In this process, helium of mass 3 is formed and neutrons are ejected. I have mentioned this particular reaction because it has the advantage of occurring at a not too high voltage (100,000 – 200,000 volts). In a not-too-distant future, the neutron source will probably be a source of heavy hydrogen ions, which are accelerated through a field of 200,000 volts and then collide with a substance containing heavy hydrogen. In fact, it must be said that sources of this type have already been made and work quite well; generally they have the defect of working for a very short time and then, for whatever reason, they break down.

The difficulties are purely technical and can be overcome. Therefore, it is reasonable to predict the possibility, even with other reactions of a similar type, of producing radioactive substances in quantities roughly equivalent to those used, for example, in medicine. And so, I am almost done. In summary, two rather different types of problems remain open. On the one hand, there are studies about the practical applications of artificial radioactivity, aiming primarily at increasing the intensity of neutron sources and, therefore, at producing considerable quantities of radioactive substances. On the other hand, remains to be studied the phenomenon itself, in order to clarify its essential characteristics.

\section*{Acknowledgements}
We would like to thank suor Virna Pasinetti for giving the paper a critical reading and for providing several helpful comments.


\begin{thebibliography}{1}



\bibitem{Nobel:1938:outreach}
Nobel Prize Outreach, ``Enrico Fermi, Prize motivation'', 1938.

\bibitem{Guerra:2005:quaderno}
F.~Guerra, N.~Robotti, ``Enrico Fermi e il quaderno ritrovato. 20 marzo 1934. La vera storia della scoperta della radioattività indotta da neutroni''
Società Italiana di Fisica, Bologna, 2015.

\bibitem{FermiE:1934:radioattivita}
E.~Fermi, ``Radioattività indotta da bombardamento di neutroni'', in
{\em La Ricerca scientifica}, vol.~1, n.~5, p.~283, 1934.

\bibitem{Holton:2005:tables}
G.~Holton, ``Enrico Fermi and the Miracle of the Two Tables'', 
in {\em Victory and Vexation in Science: Einstein, Bohr, Heisenberg, and Others},
pp.~48--64, Harvard University Press, 2005.

\bibitem{Chandrasekhar:1962:how}
S.~Chandrasekhar, note VIII, pp.~926--927 of {\em Collected Papers: (Note E Memorie)}, vol.~2.
\newblock Accademia Nazionale dei Lincei \& University of Chicago Press, 1965.

\bibitem{Segre:1970:physicist}
E.~Segrè,
{\em Enrico Fermi Physicist}, p. 80,
The University of Chicago Press, Chicago and London, 2005

\bibitem{Fermi:1935:recenti}
E.~Fermi, ``Recenti risultati della radioattivita artificiale'',
in {\em La Ricerca scientifica}, vol.~2, n.~6, pp.~399--402, 1935.
Online \url{http://digitale.bnc.roma.sbn.it/tecadigitale/giornale/TO00193681/1935/V.2/00000431}

\bibitem{Fermi:1935:recents}
E.~Fermi, ``Recent results in the study of the artificial radioactivity'',
translated into English by Leon Marshak (?) from {\em La Ricerca scientifica, vol. 2, n. 6} and re-issued as a document of the Atomic Energy Commission (NP-2052;  AEC-tr-927), 1935(?).

\bibitem{Fermi:1962:collected1}
E.~Amaldi, H.~L. Anderson, E.~Persico, E.~Segr{\'e}, and A.~Wattenberg, eds.,
{\em Enrico Fermi: Collected Papers, 1921--1938}, vol.~1 of {\em Collected Papers: (Note E Memorie)}.
\newblock Accademia Nazionale dei Lincei \& University of Chicago Press, 1962.


\end{thebibliography}
\end{document}